\documentclass[aps,onecolumn,floatfix,superscriptaddress,preprintnumbers,showpacs,showkeys]{revtex4}
\usepackage{amssymb}
\usepackage{amsmath}
\usepackage{amsfonts}
\usepackage{latexsym,graphicx,epsfig}
\usepackage{color} 
\usepackage{graphicx}

\newcommand{\eq}{\begin{eqnarray}}
\newcommand{\en}{\end{eqnarray}}

\begin{document}

\preprint{SLAC-PUB-16809}

\title{QCD Compositeness as Revealed in Exclusive Vector Boson Reactions \\ 
through Double-Photon Annihilation:
$e^+ e^- \to \gamma \gamma^\ast \to \gamma V^0 $ and 
$e^+ e^- \to \gamma^\ast \gamma^\ast \to V^0 V^0$ } 

\author{Stanley J. Brodsky} 
\affiliation{SLAC National Accelerator Laboratory, Stanford University, Stanford, CA 94309, USA} 

\author{Richard F. Lebed} 
\affiliation{Department of Physics, Arizona State University, Tempe, Arizona 85287-1504, USA} 

\author{Valery E. Lyubovitskij}
\affiliation{Institut f\"ur Theoretische Physik, Universit\"at T\"ubingen,
Kepler Center for Astro and Particle Physics, 
Auf der Morgenstelle 14, D-72076, T\"ubingen, Germany}
\affiliation{Department of Physics, Tomsk State University,  
634050 Tomsk, Russia} 
\affiliation{Laboratory of Particle Physics, 
Mathematical Physics Department, 
Tomsk Polytechnic University, 
Lenin Avenue 30, 634050 Tomsk, Russia}

\begin{abstract}

  We study the exclusive double-photon annihilation processes, $e^+
  e^- \to \gamma \gamma^\ast\to \gamma V^0$ and $e^+ e^- \to
  \gamma^\ast \gamma^\ast \to V^0_a V^0_b,$ where the $V^0_i$ is a
  neutral vector meson produced in the forward kinematical region: $s
  \gg -t$ and $-t \gg \Lambda_{\rm QCD}^2$.  We show how the
  differential cross sections $\frac{d\sigma}{dt}$, as predicted by
  QCD, have additional falloff in the momentum transfer squared $t$
  due to the QCD compositeness of the hadrons, consistent with the
  leading-twist fixed-$\theta_{\rm CM} $ scaling laws, 
  both in terms of conventional Feynman diagrams and 
  by using the AdS/QCD holographic model to obtain the results more
  transparently. However, even though they are exclusive channels
  and not associated with the conventional electron-positron
  annihilation process $e^+ e^- \to \gamma^\ast \to q \bar q,$ these
  total cross sections $\sigma(e^+ e^- \to \gamma V^0) $ and
  $\sigma(e^+ e^- \to V^0_a V^0_b),$ integrated over the dominant
  forward- and backward-$\theta_{\rm CM}$ angular domains, scale as
  $1/s$, and thus contribute to the leading-twist scaling behavior of
  the ratio $R_{e^+ e^-}$.  We generalize these results to exclusive
  double-electroweak vector-boson annihilation processes accompanied
  by the forward production of hadrons, such as $e^+ e^- \to Z^0 V^0$
  and $e^+ e^- \to W^-\rho^+$.  These results can also be applied to
  the exclusive production of exotic hadrons such as tetraquarks,
  where the cross-section scaling behavior can reveal their multiquark
  nature.

\end{abstract}



\today

\maketitle

\section{Introduction}

A surprising result, shown by Davier, Peskin, and Snyder
(DPS)~\cite{Davier:2006fu}, is that there are exclusive hadronic
contributions to the electron-positron annihilation cross section
ratio $R_{e^+e^-} = {\sigma(e^+ e^- \to X)/ \sigma(e^+ e^- \to \mu^+
  \mu^-)}$ that are scale invariant, but are not associated with the
annihilation process $e^+ e^- \to \gamma^\ast \to q \bar q.$  These
exclusive processes are based on double-photon annihilation
subprocesses, such as $e^+ e^- \to \gamma \gamma^\ast \to \gamma V^0$
and $e^+ e^- \to \gamma^\ast \gamma^\ast \to V^0_a V^0_b,$ where the
$V^0_i$ are vector bosons such as the $\rho$ meson.  Since the
amplitude involves spin-$\frac{1}{2}$ electron exchange in the $t$ and
$u$ channels, it behaves as $s^{\frac{1}{2}}$ for $s \gg -t, -u$.  The
total cross section (which includes a phase-space factor of $1/s^2$),
integrated over the dominant forward and backward regions, thus
behaves as $1/s$.

 \begin{figure}[ht]
 \includegraphics[width=0.5\textwidth]{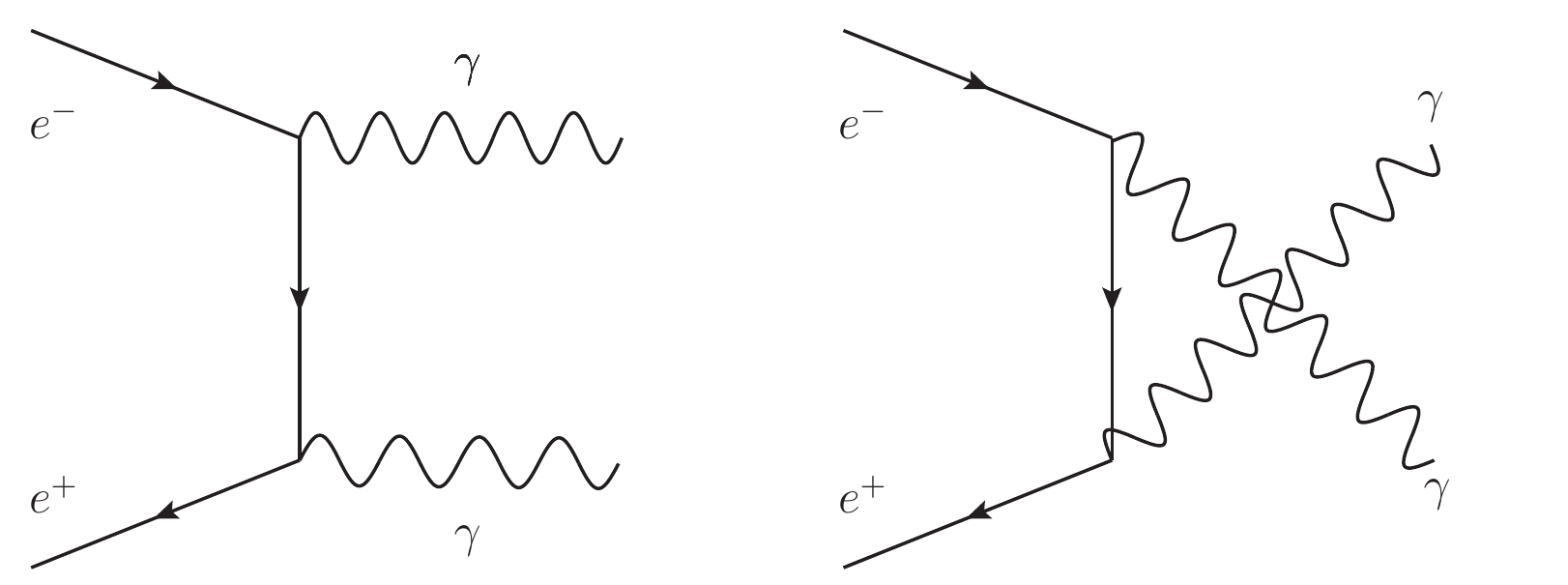}
 \caption{The double-photon annihilation amplitude in the Born
   approximation.}
          \label{fig1}
\end{figure}

 \begin{figure}[ht]
 \includegraphics[width=0.3\textwidth]{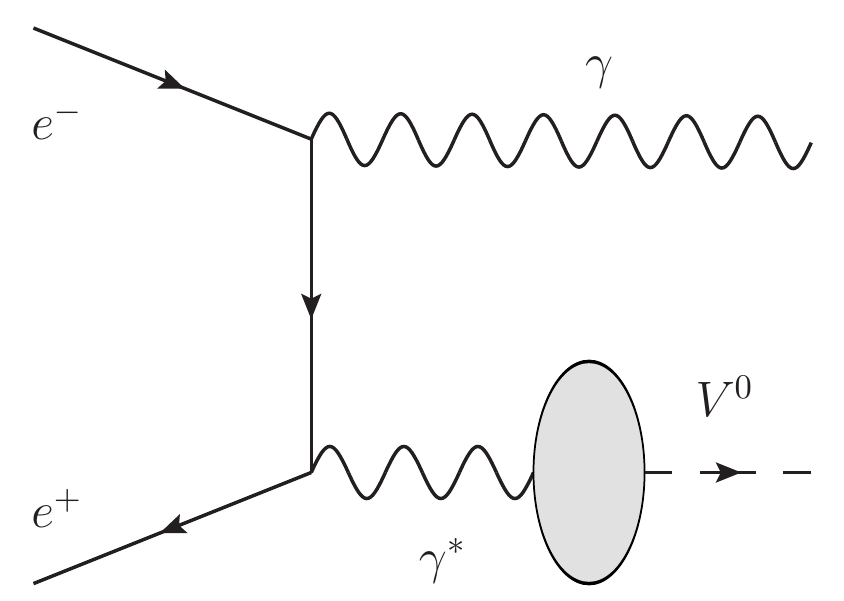}
 \caption{Exclusive production of a photon and vector boson via
   double-photon annihilation (the corresponding $u$-channel diagram
   is implied).  The differential cross section is peaked in the
   forward and backward directions.  Compositeness of the vector boson
   produces a monopole falloff of the differential cross section
   $d\sigma/dt$ in $|t|$.}
          \label{fig2}
\end{figure}

 \begin{figure}[ht]
 \includegraphics[width=0.3\textwidth]{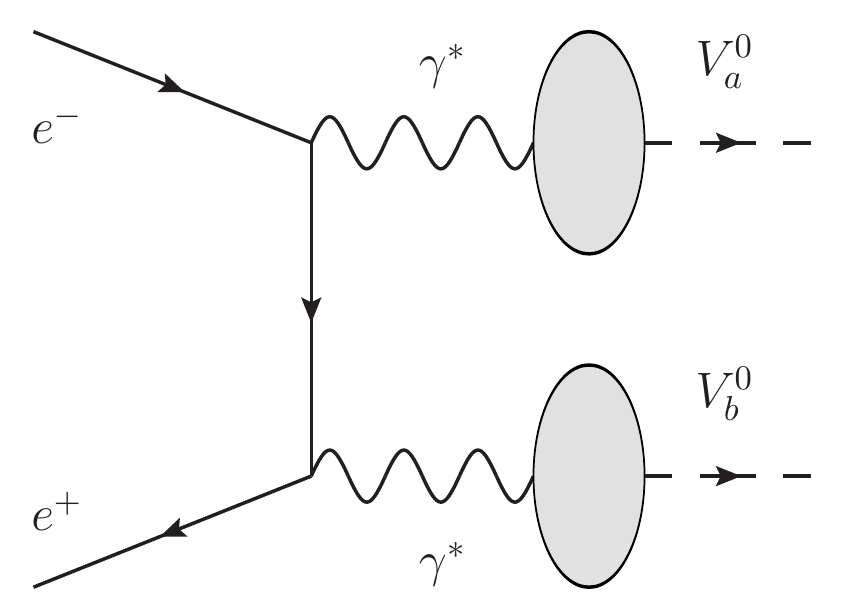}
 \caption{Exclusive production of two vector bosons via double-virtual
   photon annihilation (the corresponding $u$-channel diagram is
   implied).  Compositeness of the vector bosons produces a dipole
   falloff of the differential cross section $d\sigma/dt$ in $|t|$.}
          \label{fig3}
\end{figure}

In this paper we show how the QCD compositeness of the vector bosons
affects the matrix elements and cross sections for these double-photon
processes.  The effects of compositeness reflect the fact that the
coupling of the virtual photon proceeds through the vertex
$\gamma^\ast \to q \bar q$. One may study the scaling
behavior solely using conventional Feynman diagram techniques, as we
describe below and in Sec.~\ref{sec:coupling}.  In order to obtain
explicit closed-form results that manifest the correct scaling
behavior, we start by employing the light-front quantization of QCD
(LF-QCD); in those terms, the virtual $q \bar q$ then couples to the
valence hadronic light-front wave function $\psi_{V^0} (x, \vec
k_\perp)$.  The integration over the light-front momentum fractions
$x=k^+/P^+$ and $1-x$, and relative transverse momentum $k_\perp$, of
the pair leads to an extra factor of the QCD mass scale $\Lambda_{\rm
QCD}$ in the numerator of the amplitude. 
In order to track both the small- and large-momentum
behavior of these processes, we utilize the AdS/QCD (AdS =
anti-de~Sitter) holographic light-front
model~\cite{Brodsky:2007hb,Branz:2010ub} which is successful in
explaining the main features of meson and baryon spectroscopy, as well
as the dynamical properties of hadrons.  This nonpertubative approach
to hadron physics gives a good overall description of meson and baryon
form factors, including consistency with the perturbative QCD (pQCD)
power-law scaling of hadronic form factors at large momentum transfer.
The AdS/QCD hadronic scale $\kappa$ can be related to the slope of the
Regge trajectories, as well as providing mass relations such as
$\kappa = M_\rho/\sqrt{2}$~\cite{Brodsky:2007hb,Branz:2010ub}.  We
also show that the AdS/QCD prediction for $f_\rho$, the leptonic decay
constant of the $\rho$ meson, is in excellent agreement with
measurement. 

The double-photon $e^+ e^- \to \gamma \gamma$ amplitude illustrated in
the first inset of Fig.~\ref{fig1} behaves for large energy as
\eq 
{\cal M}(s,t) \propto \alpha_{\rm em} 
 \Big(\frac{s}{-t}\Big)^{\alpha_R}  
= \alpha_{\rm em} \Big(\frac{s}{- t}\Big)^{1/2} \, ,
\en
for $s \gg - t$, corresponding to spin-$\frac{1}{2}$ exchange in the
$t$ channel, where the differential cross section is ${d\sigma/dt}
\propto {|{\cal M}(s,t)|^2/s^2}$.  The fermion-exchange amplitudes
have both $t$- and $u$-channel contributions; however, the
interference is suppressed in the dominant forward- and
backward-peaked domains.  As we shall show, and consistent with
dimensional analysis, the differential cross section for the
production of a single vector meson $\frac{d\sigma}{dt}( e^+e^- \to
\gamma V^0)$ via double-photon annihilation (see Fig.~\ref{fig2}) must
have the extra falloff $G_V^2(t)\sim {\kappa^2/|t|}$ at large $-t \gg
\kappa^2$; {\it i.e.},
\eq
\frac{d\sigma}{dt}(e^+ e^- \to \gamma V^0) \sim 
{\frac{\alpha_{\rm em}^3}{s |t|}} {\frac{\kappa^2}{|t|}}\,.
\en
The mass parameter $\kappa$ 
is specifically the scale parameter of AdS/QCD approach; however, the
power scaling of AdS/QCD and pQCD for $G_V^2(t)$ at large $t$ are the
same, consistent with the twist dimension dictated by QCD
compositeness. Physically, the extra falloff in $|t|$ results from
the phase-space hadronization of the virtual $q \bar q$ in the
amplitude $e^+ e^- \to
\gamma q \bar q \to \gamma V^0$, which is represented by the
transition form factor $G_V(q^2)$.  In the case where two vector
bosons are produced with opposite transverse momenta (see
Fig.~\ref{fig3}), the amplitude is suppressed by two form factors, so
the differential cross section at $s \gg -t \gg \kappa^2$ scales as
\eq 
{\frac{d\sigma}{dt}}(e^+ e^- \to \gamma^\ast \gamma^\ast \to V^0_a
V^0_b)  \sim {\frac{\alpha_{\rm em}^4}{s |t|}} {\frac{\kappa^4}{t^2}}
\,.
\en
The powers of $\alpha_{\rm em}$ correspond to the couplings of the
virtual photons to the currents of the annihilating leptons and the
vector bosons.  In effect, the cross sections are the same as that
given by the naive vector-meson dominance (VMD)
model~\cite{Sakurai:1960ju} but multiplied by the form factors
required by QCD compositeness. It is worth noting
that such nontrivial form factors also naturally arise in chiral
perturbation theory calculations~\cite{Ecker:1988te,Ecker:1989yg}, but
carry a different scaling, as discussed below. 
 
The scaling results for the exclusive cross sections are consistent
with the leading-twist quark fixed-angle counting
rules~\cite{Brodsky:1973kr,Matveev:1973ra,Lepage:1980fj} :
${\frac{d\sigma}{dt}}(A+ B \to C +D) \propto F(\theta_{\rm CM} )/
s^{N-2}$, where $N= N_A + N_B + N_C + N_D $ is the total twist or
number of elementary constituents.  In our case, $N-2 =3$ for $e^+ e^-
\to \gamma V^0$ and $N-2 = 4$ for $e^+ e^- \to V^0_a V^0_b$, which
would give the scaling for non-forward angles (where $s$, $-t$, $-u$
are all of comparable size).  In the present case, the integration
over the forward peaks in $t$ and $u$ does not modify the $1/s$
scaling of the total cross section; {\it e.g.}, $\sigma_{e^+ e^- \to
  \gamma V^0}(s) \propto {\alpha^3_{\rm em}}/{s}$, up to logarithms in
$t$ (or $u$), which are cut off by the mass scales in the process:
$m_e^2$ from the propagator between the photons and $\kappa^2$ from
the dominant part of the hadronization integral.

Although compositeness does not affect the leading $1/s$ scaling of
the total cross sections of these reactions, it does strongly modify
the $t$ and $u$ dependence of the amplitudes in terms of new types of
transition form factors.  An analysis such as that given in
Ref.~\cite{Guo:2016fqg}, based on an effective field theory in which
the vector mesons are treated as elementary fields, cannot yield the
form factors and counting rules predicted by QCD due to meson
compositeness.

In addition, QCD also predicts the $\frac{1}{M^2_{V^0_i}}$ falloff of
the amplitudes as the mass of each vector boson is increased; this
falloff corresponds to the timelike $q^2$ of the virtual photon.  One
thus finds new tests of color confinement and the nonperturbative
hadronic wave functions of hadrons.  We also show that these results
can be extended to electroweak exclusive processes involving electron
or neutrino exchange, such as $e^+ e^- \to Z^0 V^0$, which are
accessible at the proposed International Lepton Collider (ILC), and to
exotic multi-quark hadrons.

\section{Coupling of Virtual Photons to Vector Bosons} \label{sec:coupling}

In this section we demonstrate  how the correct 
momentum dependence (consistent with QCD compositeness) of the
$\gamma^* \to V^0$ transition amplitude can be evaluated in the
AdS/QCD approach. The full $q^2$ dependence of the corresponding
transition form factor is model dependent, but at large values of
$q^2$, its scaling must be consistent with pQCD. 
 
First, we point out that the full off-shell $\gamma^\ast(q) \to
V^{0\ast}(q)$ transition is given by the amplitude $G_V(q^2)
(g^{\mu\nu} q^2 - q^\mu q^\nu)/q^2$, where the tensor guarantees gauge
invariance, the $1/q^2$ comes from the virtual photon propagator, and
the form factor $G_V(q^2)$ has the falloff $1/\sqrt{q^2}$ at
large~$q^2$.  Therefore, the $\gamma^\ast(q) \to V^{0\ast}(q)$
transition scales as ${\cal O}(1/\sqrt{q^2})$ at large~$q^2$.  This
scaling is interesting---it says that the couplings of heavier vector
mesons $q^2 = m_V^2$ become increasingly suppressed---but it is not
the primary scaling of interest in this paper.  
In an effective theory 
that treats the $\rho$ meson as an elementary field, the form factor
$G_V(q^2)$ is a constant.  In addition, the contributions of the
relevant diagrams are different in VMD and pQCD. For example, in
the case of the pion electromagnetic form factor, in VMD the contact
diagram gives 1 (a constant contribution at large $q^2$), whereas the
vector-meson exchange diagram gives a $(-1 + M_V^2/q^2)$ contribution.
Summing these two diagrams, we arrive at a $M_V^2/t$ scaling.  In
contrast, using pQCD counting rules, the contact diagram turns out to
be at leading order ($1/q^2$), whereas the vector-meson exchange
diagram is subleading [$(1/q^2)^{3/2}$] at large $q^2$. 

The calculation of the transition form factor $G_V(q^2)$ can be
performed in soft-wall AdS/QCD~\cite{Karch:2006pv}.  We begin by
proposing an effective action describing the coupling of the stress
tensors of two vector fields $F_{MN}$ and $V_{MN}$, which are
holographically equivalent to the electromagnetic and neutral vector
meson fields, respectively:
\eq S = - \frac 1 2 \int d^4x \, dz \, g^{(\kappa)}_{V} (z) 
e^{-\kappa^2 z^2} F_{MN}(x,z) V^{MN}(x,z) \, , \label{eq:action}
\en
where $g^{(\kappa)}_{V} (z) \equiv \frac{2 \kappa}{\sqrt{\pi}} \, g_V
z^0$ is the 5-dimensional coupling constant (which, in this special
case, is independent of $z$) and $\kappa$ is the dilaton parameter.
As seen below, $g_V$ arises as the value of $G_V(q^2)$ at $q^2 = 0$.
 
The matrix element of the
$\gamma^\ast(q) \to V^{0\ast}(q)$ transition is given by
\eq \label{eq:transition}
{\cal M}_{\rm inv}^{\mu\nu}(q) = -ie \, G_V (q^2)
(q^2 g^{\mu\nu} - q^\mu q^\nu) \,,
\en
where $G_V(q^2)$ is the transition form factor, which in the Euclidean
region is given by
\eq \label{eq:FVexact}
G_V(Q^2) =
\int\limits_0^\infty dz \, e^{-\kappa^2 z^2} g^{(\kappa)}_{V} (z)
V^2(Q,z)\,,
\en
where, in terms of the Kummer (confluent hypergeometric) function $U$,
\eq \label{eq:Vexact}
V(Q,z) 
= \kappa^2 z^2 \int\limits_0^1 \frac{dx}{(1-x)^2} x^{Q^2/(4\kappa^2)} 
e^{-\kappa^2 z^2 x/(1-x)} = \Gamma(1+a) \, U(a,0,\kappa^2 z^2) 
\en
is the bulk-to-boundary propagator for both the $F_{MN}$ and $V_{MN}$
fields~\cite{Brodsky:2007hb,Branz:2010ub}, and $a=Q^2/(4\kappa^2)$.
$V(Q,z)$ obeys the following normalization conditions:
\eq
V(0,z) = 1\,, \quad 
V(Q,0) = 1\,, \quad 
V(Q,\infty) = 0\,. 
\en 
From the first of these conditions and Eq.~(\ref{eq:FVexact}), one
easily sees that $G_V(0) = g_V$.  As $Q^2 \to \infty$, the
vector-field bulk-to-boundary propagator behaves as
\eq 
V(Q,z) & \to & \frac{Q^2z^2}{4} \, \int\limits_0^\infty \frac{dt}{t^2} \, 
e^{\kappa^2 z^2 -t- Q^2z^2 \! /(4t)} \nonumber\\
&=& e^{\kappa^2 z^2} Qz \, K_1(Qz)\,,
\en
where $K_1(x)$ is the modified Bessel function of the second kind, which 
for arbitrary $n$ is given by the integral representation 
\eq 
K_n(x) = \frac{x^n}{2^{n+1}} \, 
\int\limits_0^\infty \frac{dt}{t^{n+1}} \, 
e^{-t-x^2 \! /(4t)} \,. 
\en 
One can see that in the limit $\kappa \to 0$, the bulk-to-boundary
propagator $V(Q,z)$ in the soft-wall model coincides with the one in
the hard-wall model~\cite{Brodsky:2010ur}.  This expression for
$V(Q,z)$ gives the scaling of $G_V(Q^2) \sim 1/\sqrt{Q^2}$ for large
$Q^2$, consistent with quark-counting
rules~\cite{Brodsky:1973kr,Matveev:1973ra,Lepage:1980fj}:
\eq\label{asymptotics} 
G_V(Q^2) &=& \frac{2g_V}{\sqrt{\pi}} \, 
\frac{\kappa}{\sqrt{Q^2}}  \, 
\int\limits_0^\infty dx \, x^2 \Big[K_1(x)\Big]^2   
\, + \, {\cal O}(1/Q^2) \nonumber\\
&=& 1.044 \, g_V \,  
\frac{\kappa}{\sqrt{Q^2}}  \, 
+ \, {\cal O}(1/Q^2) \,,
\en
since
\eq 
\int\limits_0^\infty dx \, x^2 \Big[K_1(x)\Big]^2  = \frac{3\pi^2}{32}
\simeq 0.925 \,.
\en 
Had the coupling $g_V^{\kappa}(z)$ below Eq.~(\ref{eq:action}) been
chosen to scale as $z^n$, a similar calculation would produce the
form-factor scaling $1/(Q^2)^{(n+1)/2}$.  Our choice $n=0$ is the
unique one providing a bulk-independent transition between the photon
and $V$ mesons, which one expects in AdS/QCD since it mixes with the
$z$-independent kinetic terms $F_{MN} F^{MN}$ and $V_{MN} V^{MN}$.

Note the difference of our approach and the VMD model 
using an effective elementary field for the vector mesons  
is that we take into account the $Q^2$ 
dependence $G_V(Q^2)$, which leads to an additional suppression due to
$\rho$-meson compositeness; in contrast, in the VMD model, this coupling is
just a constant.  To model this result, the large-$Q^2$ dependence of
the $G_V(Q^2)$ transition form factor calculated in a soft-wall
AdS/QCD model can be approximated by the form
\eq\label{eq:ff_largeQ2}
\frac{G_V(Q^2)}{G_V(0)} = \frac{1}{1 + \sqrt{Q^2}/\Lambda_V}\,,
\en  
where $\Lambda_V = 1.044 \, \kappa \simeq \kappa$ is the scale
parameter.  In the numerical evaluation, we use $\kappa \simeq 0.5$
GeV and get $\Lambda_V = 0.522$~GeV.  This value follows from the
slopes of the hadron Regge trajectories and from the matching of the
soft-wall potential with the lattice-QCD heavy-quark potential,
leading to $\kappa^2 \simeq \sigma$, where $\sigma$ is the string
tension~\cite{Trawinski:2014msa,Gutsche:2014oua}.

\begin{figure}[ht] \includegraphics[width=0.5\textwidth]{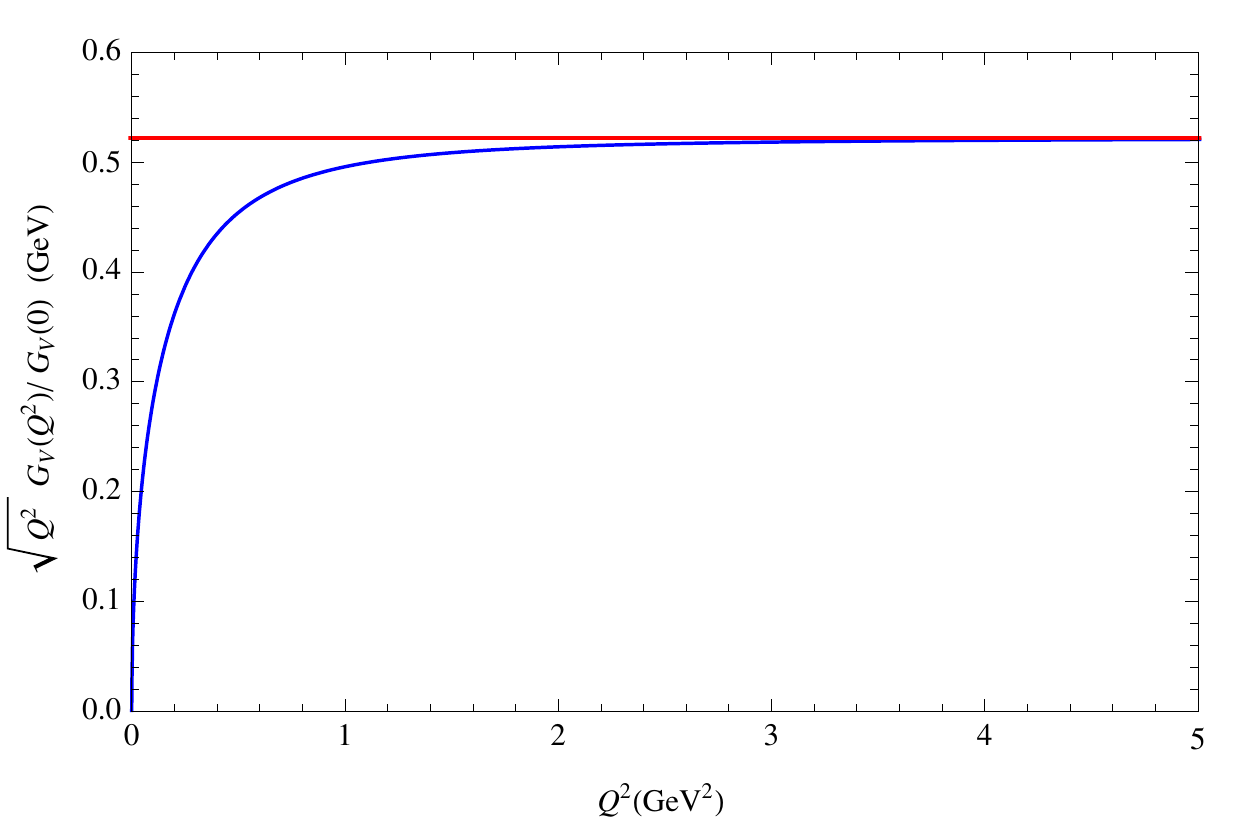}
 \caption{The form factor $\sqrt{Q^2} G_V(Q^2)/G_V(0)$. The horizontal line
 $1.044 \kappa$ corresponds to its asymptotic value at large $Q^2$.}
 \label{fig4}
\end{figure}

For the case of the $\rho$ meson, the value $G_\rho(0) = g_\rho =
0.048$ is fixed from the condition that $G_\rho(Q^2)$ determined from
Eqs.~(\ref{eq:FVexact}) and (\ref{eq:Vexact}) at the vector-meson mass
shell $Q^2 = -t = - M_\rho^2$, so that 
$F_\rho \equiv G_\rho(-M_\rho^2) = 0.202$.  
$F_\rho$, in turn, is the so-called leptonic
decay constant fixed from the central value of the data on the 
$\rho^0 \to e^+ e^-$ decay width, and is given by the
formula~\cite{Agashe:2014kda}:
\eq 
\Gamma(\rho^0 \to e^+ e^-) = \frac{4\pi}{3} \alpha_{\rm em}^2 F_\rho^2
M_\rho \,.
\en
The form factor $G_V(Q^2)$ very rapidly reaches its asymptotic value.
For example, for $\kappa = 0.5$~GeV, the idealized form of
Eq.~(\ref{eq:ff_largeQ2}) reaches about 80\% of its asymptotic value
$1.044 \, g_V \kappa/\sqrt{Q^2}$ already by $Q^2 = 4$~GeV$^2$.  The
approach to the asymptote for the exact result is even faster.  In
Fig.~\ref{fig4} we plot the result for the exact form factor
$\sqrt{Q^2} G_V(Q^2)/G_V(0)$ obtained from Eqs.~(\ref{eq:FVexact}) and
(\ref{eq:Vexact}) and compare it with asymptotic line $1.044 \,
\kappa$ deduced from Eq.~(\ref{asymptotics}).

On the other hand, the value of the coupling $F_\rho$ can be
calculated in LF-QCD, using~\cite{Branz:2010ub}           
\eq 
F_\rho = 2 \frac{\sqrt{N_c}}{M_\rho} \int\limits_0^1 dx \int \frac{d^2k_\perp}
{16 \pi^3} \, \psi_L(x,k_\perp) \,, 
\en
where $\psi_L(x,k_\perp)$ is the longitudinal light-front wave
function (LFWF) of the $\rho$ meson, and $N_c = 3$.  It is known from
the matching of LF-QCD to soft-wall AdS/QCD that~\cite{Brodsky:2011xx}
\eq 
\psi_L(x,k_\perp) = \frac{4\pi}{\kappa} \,
\frac{\sqrt{\log(1/x)}}{1-x} \, 
\exp\biggl[-\frac{k_\perp^2}{2\kappa^2}
\frac{\log(1/x)}{(1-x)^2}\biggr] \,. 
\en
Straightforward calculation gives 
\eq 
F_\rho= \frac{\kappa}{M_\rho} \, 
\sqrt{\frac{N_c}{\pi}} \, \Big(1 - \frac{1}{\sqrt{2}} \Big)\,.
\en
Using the value $M_\rho = 755.26$~MeV, one obtains from this process
that $F_\rho \simeq 0.185$, which compares well with the experimental
value $F_\rho = 0.202$ given above. Note that using the prediction of
the AdS/QCD for the $\rho$ meson mass $M_\rho = \kappa \,
\sqrt{2}$~\cite{Brodsky:2007hb,Branz:2010ub} gives startling agreement
of the calculated value of $F_\rho$ with data:
\eq
F_\rho= 
\sqrt{\frac{N_c}{2\pi}} \, \Big(1 - \frac{1}{\sqrt{2}} \Big) \simeq 0.202 \,.
\en 

\begin{figure}[ht]
 \includegraphics[width=0.3\textwidth]{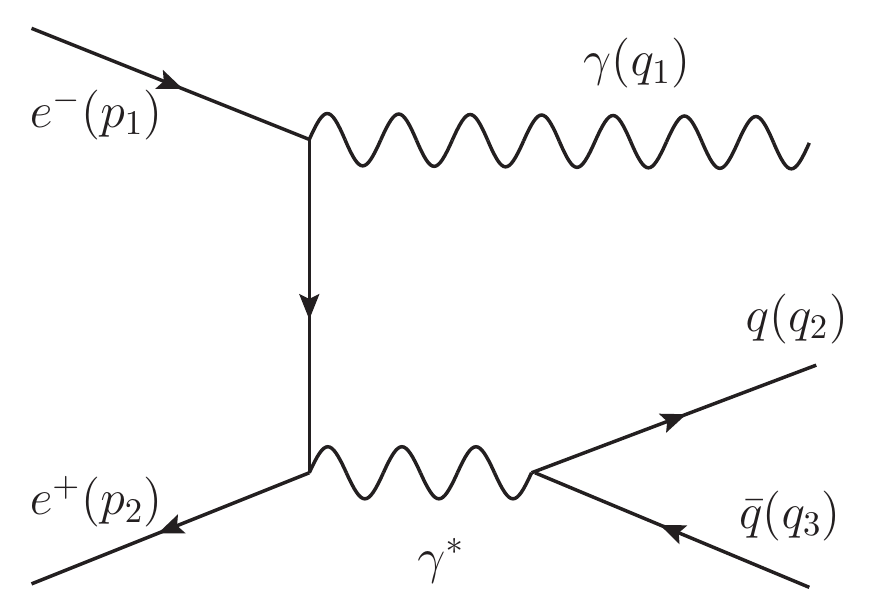}
 \caption{Diagram contributing to the $e^+ e^- \to \gamma \gamma^\ast
   \to \gamma q \bar q$ annihilation (the corresponding $u$-channel
   diagram is implied).}
          \label{fig5}
\end{figure}

Note that the $\gamma^\ast(q) \to V^{0\ast}(q)$ transition itself 
[without the form factor $G_V$] has no additional falloff in $q^2$,
because the $q^2$ from the Lorentz structure of
Eq.~(\ref{eq:transition}) is compensated by the $1/q^2$ from the
virtual photon.  This result is consistent with the VMD model, which
is based on the constant behavior of the $\gamma^\ast(q)
\to V(q)$ transition.  The main difference from VMD model is that 
QCD compositeness gives a nontrivial $t$
dependence, as discussed below.  In particular, let us evaluate the
matrix element for the process $e^+ e^- \to \gamma \gamma^\ast \to
\gamma V^0$; it is effectively given by the product of the matrix
elements of the perturbative process $e^+ e^- \to \gamma \gamma^\ast
\to \gamma q \bar q$ and the nonperturbative process $\bar q q \to
V^0$:
\eq 
{\cal M}(e^+ e^- \to \gamma V^0) 
= {\cal M}(e^+ e^- \to \gamma q \bar q) \,  
{\cal M}(\bar q  q \to V^0) \, ,
\en
where the perturbative matrix element includes the sum of $t$
and $u$ channels (see Fig.~5): 
\eq
{\cal M}(e^+ e^- \to \gamma q \bar q) =
e^3 \bar v(p_2) \,
\biggl[
\gamma^\nu \frac{1}{\not\! k} \gamma^\mu +
\gamma^\mu \frac{1}{\not\! \tilde k} \gamma^\nu
\biggr] \, u(p_1)
\, \frac{1}{s_2} \, e_q \, \bar u(q_2) \gamma_\nu v(q_3) \,
\epsilon^\ast_\mu(q_1) \,, \label{eq:pert}
\en
with $e_q$ being the quark charge, $s + t + u = s_2$, 
$s_2 = q^2 = (q_2 + q_3)^2$, $k = p_1 - q_1$, $\tilde k = q_1 - p_2$. 
Next, using Eq.~(\ref{eq:transition}) and the transversity condition
$q \cdot \epsilon^\ast(q) = 0$ for vector particles $V^0$, we replace
the product of the final-state bilinear and ${\cal M}(\bar q q \to
V^0)$ by
\eq\label{replacement}
e_q \, \bar u(q_2) \gamma_\nu v(q_3) \times {\cal M}(\bar q  q 
\to \tilde V^0) \to \tilde G_{V}(s_2; -t) \, s_2 \,
\epsilon^{\ast}_\nu(q) \,, 
\en
where the coupling function $\tilde G_{V}(s_2; -t)$ describes the
$\gamma^\ast \to q \bar q \to \tilde V^0$ transition.  At this stage,
$\tilde G_V$ is an inclusive coupling to a neutral vector source
$\tilde V^0$ at momentum transfer $s_2$ and does not yet refer to a
single vector particle $V^0$ of a particular mass $\sqrt{s_2}$.  If
$s_2$ is not otherwise constrained, one expects it generically to be
of order $-t$, as indicated by the second argument of $\tilde G_V$.
The key point to notice is that the proper amplitude to use for a
particular process depends not only upon the momentum transfer $s_2$
entering from the virtual photon, but it is further constrained by
which distinct exclusive final state it becomes; in particular, if we
consider it to be the exclusive transition form factor $G_V(s_2)$
referring to a single vector meson $V^0$.  Generically, the $q\bar q$
pair in Fig.~\ref{fig5} emerges with a large transverse momentum and
creates a number of hadrons through fragmentation; even if $s_2$ is
taken to be as small as $M_{\rho}^2$, one can still produce up to 5
pions from the $\gamma^\ast$.  For example, in order to produce a
single vector meson exclusively, whose constituent quarks are nearly
collinear, typically requires the exchange of a hard gluon between the
$q\bar q$ pair, such that the gluon momentum transfer is of the order
of $-t$.  Taking into account this physical constraint, which arises
due to QCD compositeness of the hadron, is essential for obtaining
correct scaling of the form factor $G_V$ for large $-t$; we recognize
this fact by henceforth restricting $\tilde G_V (s_2; -t) \to
G_V(M_V^2; -t) \equiv G_V(-t)$, the transition form factor for a
single vector meson $V^0$ of mass $M_V$.

Were one to insert a hard gluon line between $q_2$ and $q_3$ in
Fig.~\ref{fig5} to form a loop, each end of the (initially) hard
quarks $q_2$ and $q_3$ would provide a spinor normalization factor
$\sim |t|^{1/4}$ and each hard quark (gluon) propagator would
contribute a factor $\sim 1/|t|^{1/2}$ ($\sim 1/|t|$).  The loop
integral naively would bring in four powers of momentum, but the
leading-order piece gives rise to the divergence of the vertex
correction, which is subtracted, and the cubic piece gives an odd
function, which vanishes under integration.  Thus the loop integral
gives a contribution $\sim |t|$, so that one expects $G_V(-t)$ for
large $|t|$ to scale as $1/|t|^{1/2}$.

Finally, we obtain 
\eq
{\cal M}(e^+ e^- \to \gamma V^0) &=&
e^3 \, G_{V}(-t) \,
\bar v(p_2) \biggl[
\gamma^\nu \frac{1}{\not\! k} \gamma^\mu
+
\gamma^\mu \frac{1}{\not\! \tilde k} \gamma^\nu
\biggr] \, u(p_1)
\, \epsilon^\ast_\nu(q) \, \epsilon^\ast_\mu(q_1) \, .
\en
In comparing with Eq.~(\ref{eq:pert}) one sees that, as in the VMD
model, the denominator of the virtual photon propagator $(s_2)$ is
compensated by the factor $s_2$ coming from the Lorentz structure of
the $\bar q q \to V^0$ transition.  On the other hand, one sees the
important difference from the VMD model: Our coupling $G_{V}(-t)$ has
explicit dependence on $t$ and falls off as $1/|t|^{1/2}$ at large
$|t|$.

Our kinematical limit is as follows: We consider the small
forward-angle ($\theta$) scattering region to be such that $-t \sim s
\cdot \theta^2$, while $s_2 = M_V^2 \ll - t$ satisfies $s+t+u=s_2$. The
differential cross section $d\sigma/dt$ for the process $e^+ e^- \to
\gamma V^0$, where $V^0$ is any neutral vector-meson state, is
calculated according to the formula
\eq
\frac{d\sigma}{dt} = \frac{1}{64 \pi^2 s^2} \,
\sum\limits_{\rm pol}|{\cal M}(e^+ e^- \to \gamma V^0)|^2 \,. 
\en
After a straightforward calculation, we obtain 
\eq
\frac{d\sigma}{dt} &=& 
\frac{8\pi\alpha_{\rm em}^3}{s |t|} \, G_V^2(-t)
\biggl[ 1 + \frac{t}{s} + {\cal O}(1/s^2)  \biggr] \,. 
\en
At large $-t$, the differential cross section scales as 
\eq
\frac{d\sigma}{dt} \sim \alpha_{\rm em}^3 \, \frac{\kappa^2}{s t^2}
\,. 
\en 

For the process $e^+ e^- \to V^0_a V^0_b$, we use the following
kinematics:
\eq
s+t+u = s_{2a} + s_{2b}\,, 
\en
where $s_{2a} = q_a^2 = M_{V_a}^2$ and $s_{2a} = q_b^2 = M_{V_b}^2$,
and $s \gg - t \gg s_{2a} , \, s_{2b}$.  The differential cross
section is given by
\eq
\frac{d\sigma}{dt} &=& \frac{1}{64 \pi^2 s^2} \,
\sum\limits_{\rm pol}|{\cal M}(e^+ e^- \to V^0_a V^0_b)|^2\,, 
\en
where 
\eq
{\cal M}(e^+ e^- \to V^0_a V^0_b) &=&
e^4 \, G_{V}^2(-t) \, 
\bar v(p_2) \biggl[
\gamma^\nu \frac{1}{\not\! k} \gamma^\mu
+
\gamma^\mu \frac{1}{\not\! \tilde k} \gamma^\nu
\biggr] \, u(p_1)
\, \epsilon^\ast_\nu(q_b) \, \epsilon^\ast_\mu(q_a) \, .
\en
From this amplitude, one can compute:
\eq
\frac{d\sigma}{dt}
&=& \frac{32\pi^2\alpha_{\rm em}^4}{s |t|} \, G_V^4(-t)
\biggl[ 1 + \frac{t}{s} + {\cal O}(1/s^2)  \biggr] \,.
\en
The falloff of $\frac{d\sigma}{dt}$ for double-vector meson production
at large $t$ is thus 
\eq
\frac{d\sigma}{dt} \sim  \alpha_{\rm em}^4 \,
\frac{\kappa^4}{s |t|^3} \,.
\en
Our finding is relevant for the process of dilepton production in $
e^+ e^-$ annihilation, when the neutral vector meson produces the
lepton pair.  In this case, each vector meson gives a $1/t^2$ falloff
in the cross section, due to the falloff of the vector-meson form
factor $G_V(-t) \sim 1/|t|^{1/2}$ at large $-t$.

\section{Annihilation $e^+ e^- \to Z_c^+ \pi^-$ and 
$e^+ e^- \to Z_c^+ Z_c^-$}  

Using soft-wall AdS/QCD, we can derive the universal formula for 
the ratio
\eq 
R = \frac{\sigma(e^+ + e^- \to H_{n_1} + H_{n_2})}
{\sigma(e^+ + e^- \to \mu^+ + \mu^-)} \, ,
\en
where $n_1$ and $n_2$ are the numbers of partons in each hadron.

According to Ref.~\cite{Brodsky:2015wza}, the ratio $R$ is given by
the square of the $\gamma \to H_{n_1} + H_{n_2}$ transition form
factor:
\eq 
R = |F_{H_{n_1}H_{n_2}}(s)|^2 \, .
\en
Note that these functions $F$ are the more conventional
electromagnetic form factors (coupling two states to a photon), in
distinction to the transition form factors $G_V$ (coupling a photon to
a single state) discussed in the previous section.  In the soft-wall
AdS/QCD model, $F_{H_{n_1}H_{n_2}}(Q^2)$ is given by
\eq 
F_{H_{n_1}H_{n_2}}(Q^2) = \int\limits_0^\infty dz \, V(Q,z) 
\phi_{n_1}(z) \phi_{n_2}(z) \, ,
\en
where
\eq 
\phi_n(z) = \sqrt{\frac{2}{\Gamma(n-1)}} \, 
\kappa^{n-1} \, z^{n-3/2} \, e^{-\kappa^2z^2/2} 
\en
is the bulk profile of the AdS field dual to a hadron with $n$
constituents~\cite{Brodsky:2007hb,Gutsche:2013zia}. 
Straightforward calculation gives
\eq 
F_{H_{n_1}H_{n_2}}(Q^2) &=& \frac{\Gamma(\frac{n_1+n_2}{2}) \, 
\Gamma(\frac{n_1+n_2}{2}-1)}
{\sqrt{\Gamma(n_1-1) \Gamma(n_2-1)}} \, 
\frac{\Gamma(a+1)}{\Gamma(a+1+\frac{n_1+n_2}{2}-1)}\nonumber\\ 
&\sim& \frac{1}{a^{(n_1+n_2)/2-1}} \, ,
\en 
where $a = Q^2/(4 \kappa^2)$.  As before, the power of $z$ in the bulk
profile uniquely fixes the $Q^2$ scaling.

The latter formula reproduces the correct scaling of form factors with
the corresponding number of partons (here, $n$ is the number of $q\bar
q$ pairs):
\eq
F_{H_n} \sim \biggl(\frac{1}{Q^2}\biggr)^{n-1} \, ,
\en  
and gives a prediction for the production of single and double
tetraquarks.  In particular, the scaling of the form factor
corresponding to $\gamma^\ast \to Z_c^+ + \pi^-$ is
\eq
F_{Z_c^+\pi^-} \sim \frac{1}{Q^4} \sim \frac{1}{s^2} \, ,
\en
in case of tetraquark structure of $Z_c$ state, and 
\eq
F_{Z_c^+\pi^-} \sim \frac{1}{Q^2} \sim \frac{1}{s} \, ,
\en  
in the case when $Z_c^+$ is a system of two tightly bound diquarks.

For $\gamma^\ast \to Z_c^+ + Z_c^-$,
\eq
F_{Z_c^+Z_c^-} \sim \frac{1}{Q^6} \sim \frac{1}{s^3} \, ,
\en
in case of a $Z_c$ state with tetraquark structure, and
\eq
F_{Z_c^+Z_c^-} \sim \frac{1}{Q^2} \sim \frac{1}{s} \, ,
\en
in case when $Z_c^+$ is a system of two tightly bound diquarks.  These
results are consistent with the scaling laws discussed in
Ref.~\cite{Brodsky:2015wza} [see Eqs.~(5) and (6) in that reference].
Our results also agree with the counting rule for exclusive
tetraquark-plus-meson production discussed in
Ref.~\cite{Voloshin:2016phx}.

\section{Applications to Standard Model Processes}

\begin{figure}[ht]
\includegraphics[width=0.3\textwidth]{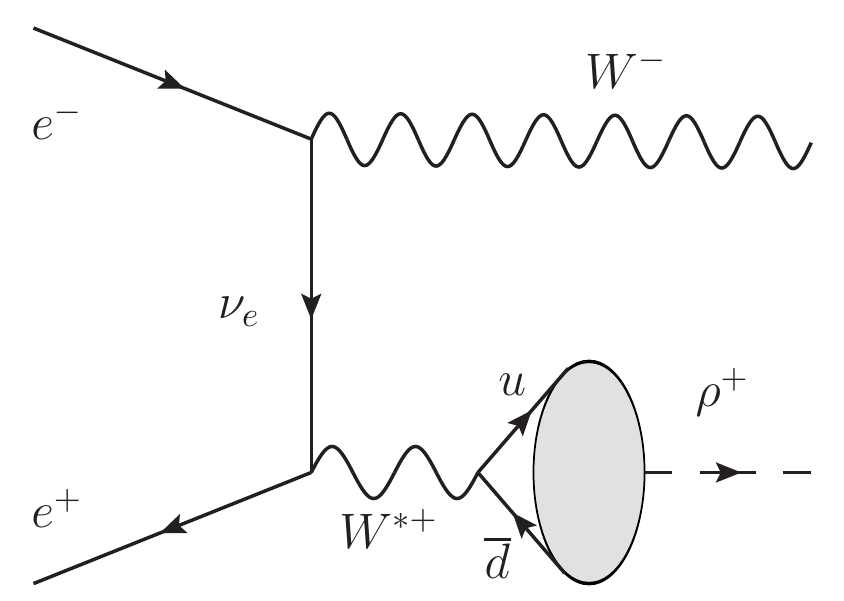}
\caption{Diagram contributing to the exclusive standard model hadronic
  amplitude $e^+ e^- \to W^{\ast +} W^- \to \rho^+ W^-$ via
  $t$-channel neutrino exchange.}  \label{fig6}
\end{figure}

We can also extend our analysis of exclusive processes to
standard-model electroweak reactions such as $e^+ e^- \to W^+ W^- \to
\rho^+ W^-$ and $e^+ e^- \to W^+ W^- \to \rho^+ \rho^-$.  An example
involving $\nu_e$ exchange is illustrated in Fig.~\ref{fig6}. The
virtual $W^+$ couples to the charged vector meson via coupling to its
$\bar u d$ valence quarks.

The cross section for $W$-pair production in $e^+ e^- $ annihilation
with one-loop corrections, along with earlier references, is given in
Ref.~\cite{Philippe:1981up}.  Neutrino exchange gives the dominant
contribution for $s \gg -t \gg M_W^2$:
\eq
\frac{d\sigma}{dt}(e^+ e^- \to W^+ W^-) \simeq
\frac{\alpha_{\rm em}^2}{4 \sin^4\theta_W} \, \frac{|t|}{4 s M_W^4} \,.
\en
In the case $s \gg M_W^2 \gg -t$, we obtain:  
\eq
\frac{d\sigma}{dt}(e^+ e^- \to W^+ W^-) \simeq
\frac{\alpha_{\rm em}^2}{4 \sin^4\theta_W} \, \frac{1}{s |t|} \,.
\en
The energy dependence for $s \gg M_W^2 \gg -t $ at fixed $t$ is
consistent with expectations for spin-$\frac{1}{2}$ exchange.

As is the case for $e^+ e^- \rightarrow \gamma \gamma^* \rightarrow
\gamma V^0$ reactions, the effect of hadron compositeness is to modify
the analytic dependence in $t$ of the exclusive cross section by a
monopole form factor at $s \gg -t \gg M_W^2$:
\eq
\frac{d\sigma}{dt}(e^+ e^- \to W^- W^{\ast +} \to W^- V^+)
\simeq
\frac{\pi \alpha_{\rm em}^3 |V_{ud}|^2}{4 \sin^6\theta_W} \,
\frac{G_V^2(-t) \, M_\rho^2 |t|}{4 s M_W^6}
\sim
\alpha_{\rm em}^3 |V_{ud}|^2 \, \frac{\kappa^2  M_\rho^2}{s M_W^6} \,.
\en 
In the case $s \gg M_W^2 \gg -t$, we find: 
\eq
\frac{d\sigma}{dt}(e^+ e^- \to W^- W^{\ast +} \to W^- V^+)
\simeq
\frac{\pi \alpha_{\rm em}^3 |V_{ud}|^2}{4 \sin^6\theta_W} \,
\frac{G_V^2(-t) \, M_\rho^2}{2 s M_W^4}
\sim
\alpha_{\rm em}^3 |V_{ud}|^2 \, \frac{\kappa^2  M_\rho^2}{s |t| M_W^4} \,.
\en

Remarkably, the differential cross section is independent of $t$ when
$s \gg -t \gg M_W^2$, and the integrated cross sections for these
exclusive hadronic reactions satisfy the leading-twist asymptotic
scaling of the form $s R_{e^+ e^-}(s) \sim {\rm const}$, despite the
compositeness of the vector mesons.

Similar results also hold for neutral vector meson reactions such as
$e^+ e^- \to Z^{\ast 0} Z^0 \to V^0 Z^0$ and $e^+ e^- \to Z^{\ast 0}
Z^{\ast 0} \to V^0 V^0$.  These processes can all be studied at the
proposed high-energy International Linear
Collider~\cite{vanderKolk:2016akp}.

\section{Conclusions}

We have seen that constituent counting rules can be used to develop
interesting predictions for a variety of production processes in the
forward and backward directions for exclusive $C=+$ annihilation
processes involving fermion exchange, such as $e^+ e^- \to \gamma V^0$
and $e^+ e^- \to V^0 V^0.$ In such cases, keeping separate track of
powers of $1/s$ and $1/|t|$ leads to interesting predictions,
depending upon the states produced.  The power-law falloff in the
Mandelstam variables due to hadron compositeness can be determined
both through the use explicit AdS/QCD soft-wall models and through
consideration of the underlying quark amplitudes.

Processes in which a meson is created by a single photon introduce a
transition form factor that must scale as $1/|t|^{1/2}$ for $-t \gg
\Lambda_{\rm QCD}^2$, so that, for example, the differential cross
section for $e^+ e^- \to \gamma \gamma$ in the forward direction
scales as $1/(s|t|)$ when $s \gg -t \gg \Lambda_{\rm QCD}$, but the
exclusive cross section for $e^+ e^- \to \gamma \rho^0$ in the forward
direction scales as $1/(st^2)$ in the same kinematical limit.
Similarly interesting results hold for high-energy electroweak
processes such as forward-angle production $e^+ e^- \to \rho^+ W^-$,
which must await the construction of the International Linear
Collider.

The consequences of the constituent counting rules have also been
verified here using AdS/QCD soft-wall models for the form factors of
multiquark exotic hadrons.  A core message of this paper is precisely
the same as that which informs much of the work using constituent
counting rules: In order to force a number of elementary constituents
created at high energies and at large relative angles into hadrons in
which they have small relative momenta, a number of
hard constituents must be exchanged between them in order to produce
bound hadronic states.  This effect will only become more pronounced
as experiments reach ever higher energies.  

\begin{acknowledgments}
  We thank Feng-Kun~Guo for an exchange that inspired us to undertake
  this work. We also thank Michael Peskin for helpful conversations.  
  This research was supported by the U.S.\ Department of
  Energy, contract DE--AC02--76SF00515 (SJB), by the U.S.\ National
  Science Foundation under Grant No.\ PHY-1403891 (RFL), by the German
  Bundesministerium f\"ur Bildung und Forschung (BMBF) under Project
  05P2015 - ALICE at High Rate (BMBF-FSP 202): ``Jet and fragmentation
  processes at ALICE and the parton structure of nuclei and structure
  of heavy hadrons'', by Tomsk State University Competitiveness
  Improvement Program, and the Russian Federation program ``Nauka''
  (Contract No.\ 0.1526.2015, 3854) (VEL). SLAC-PUB-16809.
\end{acknowledgments}.

\end{document}